# Angular dependence of the Wigner time delay upon tunnel ionization of $H_2$

D. Trabert[1] ✉, S. Brennecke[2], K. Fehre[1], N. Anders[1], A. Geyer[1], S. Grundmann[1], M. S. Schöffler[1], L. Ph. H. Schmidt[1], T. Jahnke[1], R. Dörner[1], M. Kunitski[1] & S. Eckart[1] ✉

When a very strong light field is applied to a molecule an electron can be ejected by tunneling. In order to quantify the time-resolved dynamics of this ionization process, the concept of the Wigner time delay can be used. The properties of this process can depend on the tunneling direction relative to the molecular axis. Here, we show experimental and theoretical data on the Wigner time delay for tunnel ionization of $H_2$ molecules and demonstrate its dependence on the emission direction of the electron with respect to the molecular axis. We find, that the observed changes in the Wigner time delay can be quantitatively explained by elongated/shortened travel paths of the emitted electrons, which occur due to spatial shifts of the electrons' birth positions after tunneling. Our work provides therefore an intuitive perspective towards the Wigner time delay in strong-field ionization.

[1] Institut für Kernphysik, Goethe-Universität Frankfurt am Main, Max-von-Laue-Straße 1, Frankfurt am Main 60438, Germany. [2] Institut für Theoretische Physik, Leibniz Universität Hannover, Appelstraße 2, Hannover 30167, Germany. ✉email: trabert@atom.uni-frankfurt.de; eckart@atom.uni-frankfurt.de

More than 100 years after the discovery of the photoelectric effect and its explanation in the energy domain[1], its actual duration is still heavily studied. Experiments addressing the time-resolved dynamics of the photoelectric effect for single-photon and strong-field ionization became feasible during the last 10 years[2–8]. The concept of the Wigner time delay[9] has been employed to characterize such time-resolved ionization dynamics. A missing piece, which has not been studied, so far, is the Wigner time delay for the strong-field ionization of aligned molecules. Historically, the Wigner time delay, $\tau_{\text{Wigner}}$, was introduced to describe scattering processes[9] and it is defined as the derivative of the phase of the electron's wave function $\psi$ with respect to the electron's energy $E$[5,10–13]:

$$\tau_{\text{Wigner}} := \hbar \frac{d \arg(\psi)}{dE} \quad (1)$$

According to this definition, $\tau_{\text{Wigner}}$ characterizes the spectral phase of the photoelectron wave packet and is closely related to its group delay[11–13]. In the original context of a scattering process, the origin of a phase shift is the potential (by which the electron is scattered) as it modulates the electron's wavelength upon passage. In an ionization process, however, the electron resides initially inside the potential of its parent ion and finally escapes from it. In such a "half-scattering" scenario, further parameters influence the electron's final phase shift, for example, details of the interaction process that launches the electron wave[14] and the exact location from which the wave emerges[5]. To set the scope of this paper, we express the Wigner time delay as the sum of two components:

$$\tau_{\text{Wigner}}(E, \beta) = \tau_{\text{W,A}}(E) + \Delta\tau_{\text{W,M}}(E, \beta) \quad (2)$$

In this work, we employ circularly polarized laser light for the emission of the photoelectron. The angle $\beta$ is the electron's emission angle with respect to the molecular axis within the light's polarization plane. The two contributions to $\tau_{\text{Wigner}}$ are defined such, that, for a given electron energy $E$, the average of $\Delta\tau_{\text{W,M}}(E,\beta)$ with respect to $\beta$ is zero and, $\tau_{\text{W,A}}(E)$ models the energy dependence of the Wigner time delay that is independent of $\beta$. It is important to note that, the term $\tau_{\text{W,A}}(E)$ is not addressed in our current work. The scope of this paper is to investigate $\Delta\tau_{\text{W,M}}(E,\beta)$, which quantifies the changes of the Wigner time delay as a function of the relative emission direction of the electron with respect to the molecular axis for strong-field ionization by circularly polarized light.

## Results

**The expectation for $\Delta\tau_{\text{W,M}}(E,\beta)$ based on an intuitive model.** Let us start by considering possible, intuitive reasons for the occurrence of a Wigner time delay contribution, $\Delta\tau_{\text{W,M}}(E,\beta)$, that depends on the molecular frame emission direction of the electron. Strong field ionization in the tunneling regime is often described as a two-step process. First, the electron leaves its initial bound state by tunneling and afterward, the electron follows a classical trajectory that is determined by the classical dynamics in the time-dependent potential formed by the laser field. If this classical trajectory is shifted in position space in the direction that is parallel (antiparallel) to the electron's emission direction, then the electron arrives earlier (later) at a hypothetical detector[13]. Building on this trivial idea, Fig. 1 illustrates how the initial tunneling direction with respect to the molecular axis of $H_2$ affects $\Delta\tau_{\text{W,M}}$ for strong-field ionization by circularly polarized light. We employ the partial Fourier transform (PFT) model[15] and use the linear combination of atomic orbitals (LCAO) approach to obtain the single electron wave function of each of the two electrons of $H_2$ as the sum of two $1s$ orbitals, that are separated by the molecule's internuclear distance of 0.74 Å[16] as input. The resulting electron density in the $xy$-plane $|\psi(x,y)|^2$ is shown in Fig. 1a, for a molecular axis that is aligned along the $y$-direction. The tunneling process is modeled by projecting the bound electronic wave function to the line $s_\perp$ which is perpendicular to the tunneling direction[15] (see "Methods" and refs. [15–17] for details). Due to the shape of the bound electronic wave function of $H_2$, the peak position of the one-dimensional projection changes by $\Delta s_\perp$ along $s_\perp$ as the relative angle $\gamma$ between the electron's tunneling direction and the molecular axis is altered. As we show in Fig. 1b, the displacement of the projected wave function can be quite substantial, with $\Delta s_\perp$ reaching values up to 0.13 Å for $\gamma = \pm 45°$ and the full behavior of $\Delta s_\perp$ as a function of $\gamma$ is shown in Fig. 1c. For the case of strong-field ionization employing circularly polarized laser light (the $xy$-plane is the light's polarization plane), the final electron momentum $\mathbf{p}_{\text{elec}}$ is known to be perpendicular to the initial tunnel direction if Coulomb interaction after tunneling is neglected. In this case, the electron emission angle $\beta$ in the molecular frame is given by $\beta = \gamma - 90°$ for the light-helicity that is indicated in Fig. 1. Hence, the final electron momentum is parallel or antiparallel to $s_\perp$[13,18,19] and one can relate the position-space offsets $\Delta s_\perp$ directly to changes of the arrival time at a hypothetical detector (that is far away from the electron's birth position)[13] by $\Delta\tau = \frac{\Delta s}{v} = -\frac{\Delta s_\perp m_e}{|\mathbf{p}_{\text{elec}}|}$. Figure 1d displays the corresponding emission angle-dependent delays, that can be expected from this simple modeling, which vary from $-20$ attoseconds (as) to $+20$ as. The vanishing contribution of $\Delta\tau$ for $\beta = 0°$ and $\beta = \pm 90°$ can be explained by vanishing position offsets due to the symmetry of $H_2$ and the magnitude of $\Delta\tau$ decreases for increasing electron energy $E$.

In this classical perspective, $\Delta\tau$ occurs solely due to a displacement of the mean birth location of the electron. The resulting delay is, however, consistent[13] with the definition of the $\beta$- and energy-dependent contribution to the Wigner time delay $\Delta\tau_{\text{W,M}}$ (Eqs. 1 and 2). This can be seen by referring to the basic property of quantum mechanics[20] that a translation of a wave packet in position space by $\Delta s_\perp$ is equivalent to a change of the linear phase gradient in momentum space by $\Delta\phi'_{\text{init}} = -\Delta s_\perp/\hbar$. The corresponding Wigner time-dependence is, therefore, indeed, given by $\Delta\tau_{\text{W,M}} = \frac{\hbar m_e}{|\mathbf{p}_{\text{elec}}|}\Delta\phi'_{\text{init}}$ (see "Methods" for details).

**Introduction to holographic angular streaking of electrons.** In order to determine $\Delta\tau_{\text{W,M}}(E,\beta)$ experimentally, one has to access properties of the phase of the photoelectrons' wave function in momentum space. While the absolute phase of a quantum mechanical wave function is experimentally not accessible, relative phases can be measured via interference. Many experimental techniques that address the Wigner time delay employ two or more interfering pathways. For single-photon ionization, one such scheme is the reconstruction of attosecond harmonic beating by interference of two-photon transitions (RABBITT)[21]. In RABBITT, there are two pathways that lead to the same final electron energy. On each pathway, two photons from two different laser pulses are absorbed, and the time delay between the two pulses is varied[22] (see ref. [23] for a proposed generalization of RABBITT to the multiphoton regime). The main challenge for corresponding studies with respect to strong-field tunnel ionization is that many photons are already absorbed during the ionization process. This leads to a plethora of possible pathways in the energy domain, that must be considered in order to understand the observed interference[23,24]. To meet that challenge, a recently suggested alternative interferometric approach toward $\Delta\tau_{\text{W,M}}$ in momentum space termed "holographic angular streaking of electrons" (HASE)[13] can be used which builds on refs. [25,26]. This scheme exploits semi-classical trajectories to model the interference occurring for tunnel ionization[25,27–29] triggered by a co-rotating two-color (CoRTC) laser field. According to that model, a phase gradient of the initial wave packet (which is linked to the Wigner time delay) manifests as a macroscopic rotation of a characteristic interference pattern in the electron momentum distribution. Here, we present an experiment exploiting this scheme in order to retrieve the angular dependence of the Wigner time delay $\Delta\tau_{W,M}(E,\beta)$ from the measured electron momentum distributions.

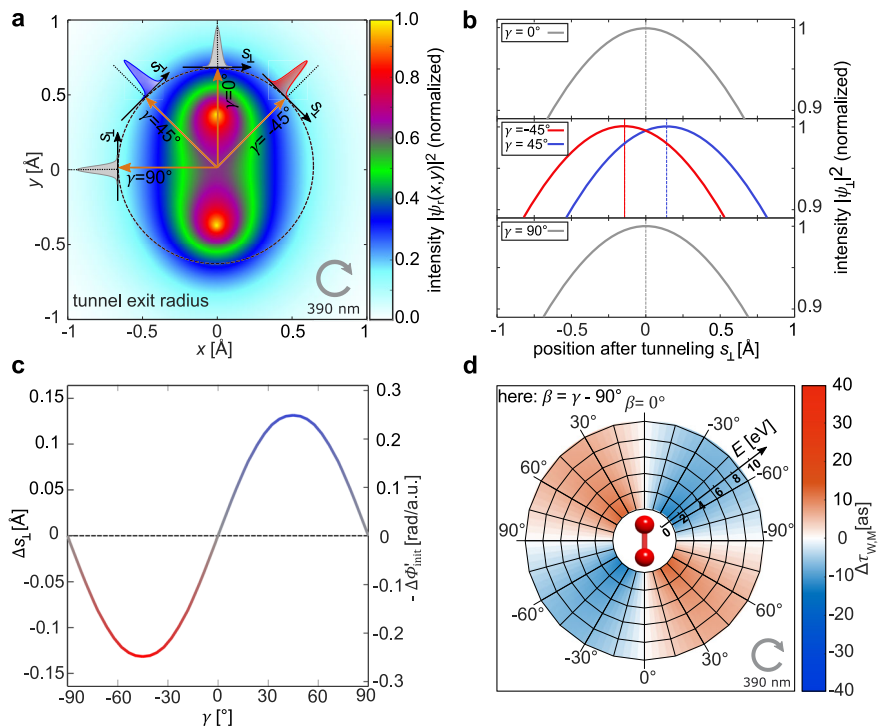

**Fig. 1 Origin of the dependence of the Wigner time delay on the molecular orientation for strong-field ionization. a** Two-dimensional distribution $|\psi_r(x,y)|^2$ of the molecular orbital of $H_2$ in position space. Different tunneling directions (orange arrows, defined by the angle $\gamma$) and the corresponding perpendicular directions (black arrows, labeled with $s_\perp$) are indicated. The colored Gaussian curves allude to the dependence of the projection of $|\psi_r(x,y)|^2$ along $s_\perp$ as a function of $\gamma$ that is discussed in (**b**, **c**). The tunnel exit position that is used in (**a**) is chosen unrealistically short (about a factor of 12) for illustrational purposes (see "Methods"). **b** One-dimensional cuts of the square of the wave function in position space along the four exemplary arrows labeled with $s_\perp$ from (**a**). The position offset $\Delta s_\perp$ is zero for tunneling parallel or perpendicularly to the molecular axis ($\gamma = 0°$ and $\gamma = \pm 90°$) due to the symmetry of $H_2$. Maximally positive [negative] values of $\Delta s_\perp$ appear for $\gamma = 45°$ [$\gamma = -45°$]. **c** The position offset $\Delta s_\perp(\gamma)$ of the one-dimensional cuts of the square of the wave function in position space along $s_\perp$. Note, that $\Delta s_\perp$ is equivalent to $-\Delta\phi'_{\text{init}}$ (see text and "Methods"). **d** The values of $\Delta\tau_{W,M}$ as a function of the electron energy $E$ and $\beta$, which is the electron emission direction with respect to the molecular axis, are obtained using the curve from (**c**) and $\Delta\tau_{W,M} = \frac{\hbar m_e}{|p_{\text{elec}}|}\Delta\phi'_{\text{init}}$ (see Methods). Note that $\beta = \gamma - 90°$ within the PFT model and that the width of the illustrated wave packets in (**a**) is not to scale.

The tailored laser electric field $\mathbf{E}(t)$, that is used in our experiment, is shown in Fig. 2a. $\mathbf{E}(t)$ is a CoRTC field that is generated by superimposing two femtosecond laser pulses: one pulse with high intensity and a central wavelength of 390 nm and one low-intensity pulse at a central wavelength of 780 nm. Both single-color fields are circularly polarized and have the same helicity. The electric field is strong enough to bend the binding potential of the atom, giving rise to a rotating barrier through which an electron can tunnel from its bound state. To qualitatively explain the idea of HASE, we neglect Coulomb interaction after tunneling. For an electron that tunnels at time $t_0$ with an initial momentum $\mathbf{p}_{\text{init}}$, the final electron momentum $\mathbf{p}_{\text{elec}}$, which is gained by the electron until the end of the laser pulse, is given by $\mathbf{p}_{\text{elec}} = \mathbf{p}_{\text{init}} - e\mathbf{A}(t_0)$. $\mathbf{A}$ is the laser's vector potential, $e$ is the elementary charge and $\mathbf{p}_{\text{init}}$ is assumed to be perpendicular to the electric field at the instant of tunneling (i.e., $\mathbf{p}_{\text{init}} \cdot \mathbf{E}(t_0) = 0$ for every trajectory)[30,31]. The two half-cycles (labeled as "c1" and "c2" in Fig. 2a) differ in their electric field,

$\mathbf{E}(t)$, as well as in their negative vector potential, $-\mathbf{A}(t)$. Consequently, there are two different combinations of $\mathbf{A}(t_0)$ and $\mathbf{p}_{\text{init}}$ within one full cycle of the laser field that lead to the same final electron momentum. These give rise to interference and allow for the retrieval of changes in the Wigner time delay. It has been shown that this description as a two-path interference also holds if Coulomb interaction after tunneling is included in the modeling[13,29]. Noteworthy, the changes of the Wigner time delay can be measured with attosecond precision, although laser pulses with durations of several dozens of femtoseconds are used.

**Holographic angular streaking of electrons in the molecular frame**. In order to measure $\Delta\tau_{W,M}$ as a function of the emission direction of the electron with respect to the molecular axis, the electron momentum vector as well as the spatial orientation of the molecule has to be measured for each ionization event. To that end, we focus the laser pulses onto a cold molecular beam of $H_2$ and detect the electron and the proton momentum for ionization and subsequent dissociation into $H^0$ and $H^+$ using a Cold Target Recoil Ion Momentum Spectroscopy (COLTRIMS) reaction microscope[32] (see "Methods"). The dissociation is much faster than the rotation of the intermediately formed $H_2^+$ molecule and thus the directions of the fragments' momenta coincide with the molecular axis at the instant of ionization[33].

Figure 2b shows the electron momentum distribution in the laser polarization plane, which exhibits an AHR pattern. This pattern in momentum space can be divided into half-rings that belong to above-threshold ionization (ATI) peaks in the electron energy spectrum and half-rings that are related to sideband (SB) peaks[26,29,34]. This definition is chosen such that the SBs vanish if the pulse with a central wavelength of 780 nm is switched off (the remaining ATI half-rings turn into full rings in this case). Effectively, for a given value of $E$ and $\beta$ the tunneling direction of the electron relative to the molecular axis is fixed. For each value of $\beta$ there is a full electron momentum distribution as in Fig. 2b that is analyzed independently which also ensures that the orientation dependence of the yield in the molecular frame[35] does not influence the conclusions drawn from the data (see Fig. S1 and "Methods").

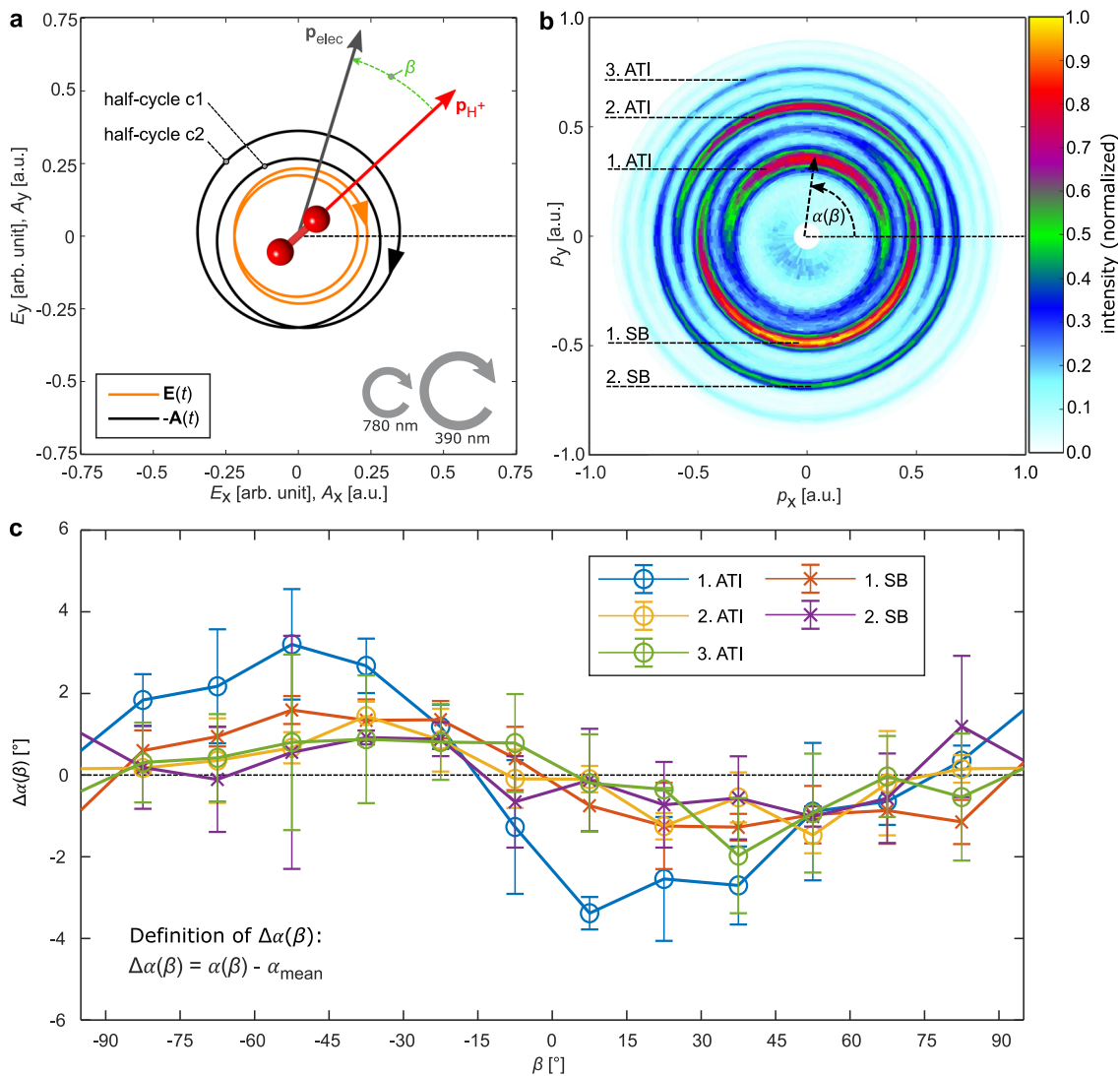

**Fig. 2 Overview of the experimental quantities that give access to changes of the Wigner time delay. a** Electric field **E**($t$) and negative vector potential $-$**A**($t$) for one cycle of the co-rotating two-color (CoRTC) field comprised of a high-intensity pulse (central wavelength of 390 nm) and a low-intensity pulse (central wavelength of 780 nm). The helicities of the two pulses are indicated with arrows. Using that the ion's momentum vector $\mathbf{p}_{H^+}$ always points along the molecular axis allows for the measurement of the orientation of the molecular axis. $\mathbf{p}_{H^+}$ is measured in coincidence with the electron momentum vector $\mathbf{p}_{elec}$. As illustrated, $\beta$ is the electron emission angle relative to the molecular axis in the polarization plane. **b** Measured electron momentum distribution in the polarization plane of the laser's electric field: ATI and SB peaks are half-rings which are spaced by the energy of a photon of the weaker laser pulse at 780 nm (1.6 eV). The most probable electron emission direction, $\alpha(\beta)$, is indicated for the first ATI peak. **c** Changes in the most probable electron angle, $\Delta\alpha$, as a function of $\beta$ are presented for each ATI/SB peak separately. $\alpha_{mean}$ is determined for every energy peak independently as the mean of $\alpha(\beta)$ over all $\beta$ (see Fig. S1). The values for the vector potential and the electron momentum are in atomic units (a.u.). The error bars show the standard deviation of the statistical errors.

As indicated in Fig. 2b, there exists a most probable electron emission angle $\alpha$ for each half-ring. The changes of $\alpha$ are investigated as a function of $\beta$. Figure 2c shows the difference

$\Delta\alpha(\beta) = \alpha(\beta) - \alpha_{mean}$, where $\alpha_{mean}$ is the most probable electron emission angle integrated over all values of $\beta$ (see "Methods" and Fig. S1). For each ATI/SB half-ring $\alpha_{mean}$ and $\Delta\alpha(\beta)$ are determined independently. It is evident that $\Delta\alpha(\beta)$ varies on the order of a few degrees and that the overall shape of the curves is similar for ATI peaks and SB peaks. A decrease in the modulation amplitude for higher electron energies is observed.

**Position offsets at the tunnel exit.** The curves showing $\Delta\alpha$ as a function of $\beta$ (Fig. 2c) contain all information that is needed to infer the changes of the phase gradient $\Delta\phi'_{init}$ as a function of the tunneling angle in the molecular frame $\gamma$ (see Fig. 1a) for each energy peak. For the quantitative analysis of the experimental data, Coulomb interaction after tunneling is taken into account by using a semiclassical two-step (SCTS) model (see "Methods" and Fig. S2). The experimentally retrieved value for $\Delta\phi'_{init}$ is shown in Fig. 3a for each energy peak. For comparison, Fig. 3b depicts the result for $\Delta\phi'_{init}$ from strong-field approximation (SFA, see "Methods") and the PFT model. Experiment and SFA are in excellent agreement and show that $\Delta\phi'_{init}$ depends on the initial momentum $p_{initial}$. Interestingly, within the PFT model, $\Delta\phi'_{init}$ is independent of $p_{initial}$. The reason for this is that the PFT model approximates the tunneling process to be adiabatic whereas SFA includes nonadiabatic tunneling[31,36,37]. From this we conclude that the experimentally observed dependence of $\Delta\phi'_{init}$ on $p_{initial}$ is a fingerprint of the nonadiabatic dynamics in the classically forbidden region (tunnel). From a position space perspective, this reveals that for nonadiabatic tunneling, the position offset $\Delta s_\perp = -\Delta\phi'_{init}/\hbar$ depends on $p_{initial}$.

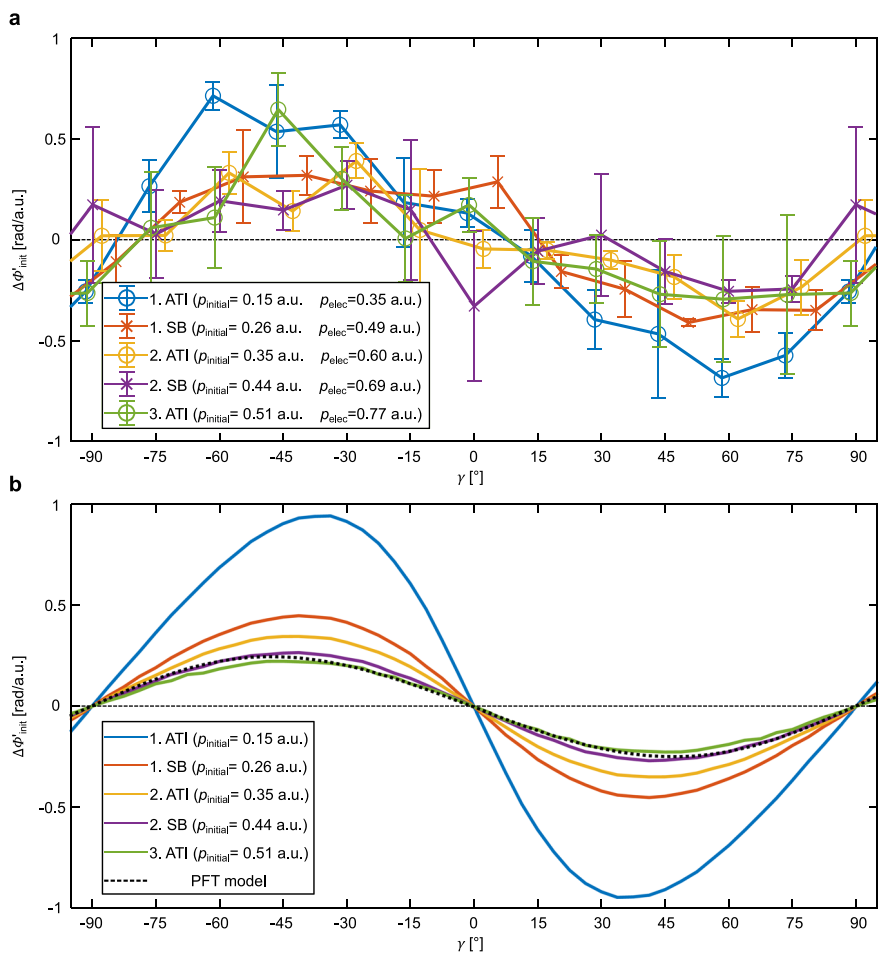

**Fig. 3 Dependence of the phase gradient at the tunnel exit on the molecular orientation. a** Experimentally retrieved phase gradient $\Delta\phi'_{init}$ at the tunnel exit as a function of $p_{initial}$ (initial momentum at the tunnel exit) and $\gamma$ (relative angle between the electron's tunneling direction and the molecular axis of $H_2$ in the polarization plane). The error bars show the standard deviation of the statistical errors. **b** Shows the theoretically obtained phase gradient $\Delta\phi'_{init}$ at the tunnel exit from strong-field approximation (SFA). The result from the PFT model is shown for comparison as a dotted black line in (**b**). The dashed black lines in (**a**, **b**) are horizontal lines to guide the eye.

**The experimentally obtained value for $\Delta\tau_{W,M}(E,\beta)$.** The experimentally obtained phase gradient $\Delta\phi'_{init}$ at the tunnel exit can be used to determine the change of the Wigner time delay $\Delta\tau_{W,M}(E,\beta)$. To this end, we use a trajectory-based classical mapping of the initial velocities to the final momenta including a long-range $1/r$ potential. This enables us to map the obtained phase gradient $\Delta\phi'_{init}$ to changes of the Wigner time delay $\Delta\tau_{W,M}$ (see "Methods"). The experimental result for $\Delta\tau_{W,M}(E,\beta)$ is shown in Fig. 4a and is in good qualitative agreement with the results from the PFT model (see Fig. 1d). The absolute magnitude of $\Delta\tau_{W,M}$ (note that the color scale in Figs. 1d and 4a is identical), the change of sign between the quadrants and the decrease of $\Delta\tau_{W,M}$ with electron energy are faithfully reproduced. This supports, that the PFT model illuminates the physics behind the $\beta$-dependence of the Wigner time delay in strong-field ionization, which is the spatial displacement of the tunneling wave packet that gives the electron a head start, or a longer way to travel.

**Comparison of experiment and theory.** To benchmark our findings using ab initio theory, we have determined $\Delta\tau_{W,M}(E,\beta)$ by solving the time-dependent Schrödinger equation (TDSE) numerically. This approach is completely independent to the procedure that is used in the experiment, since for our TDSE simulation $\Delta\tau_{W,M}$ is directly retrieved from the phase of the continuum wave function (see Methods). Comparison of Fig. 4a and 4b shows that the result from the TDSE simulation shows the same magnitudes as in the experiment and that besides a slight offset in $\beta$, the TDSE result and experiment show excellent agreement.

To further discuss our experimental findings, we show the result from SFA in Fig. 4c, where the zero crossing of $\Delta\tau_{W,M}$ is at $\gamma = 0°$ and $\gamma = \pm 90°$ as in the PFT model (see Fig. 1d). Evidently, the result from SFA shows significantly larger magnitudes of $\Delta\tau_{W,M}$ compared to the experiment and the TDSE simulation. These larger values occur because the deceleration of the electron by the long-range ionic potential is neglected in SFA. To overcome this limitation of SFA, we have used the same classical mapping as before (in full analogy to the procedure to obtain the experimental result in Fig. 4a, see Methods for details). The result is shown in Fig. 4d) and it is clearly seen that the inclusion of the long-range $1/r$ potential leads to a decrease of the magnitude of $\Delta\tau_{W,M}$ and induces a slight rotation which is in excellent agreement with the experiment.

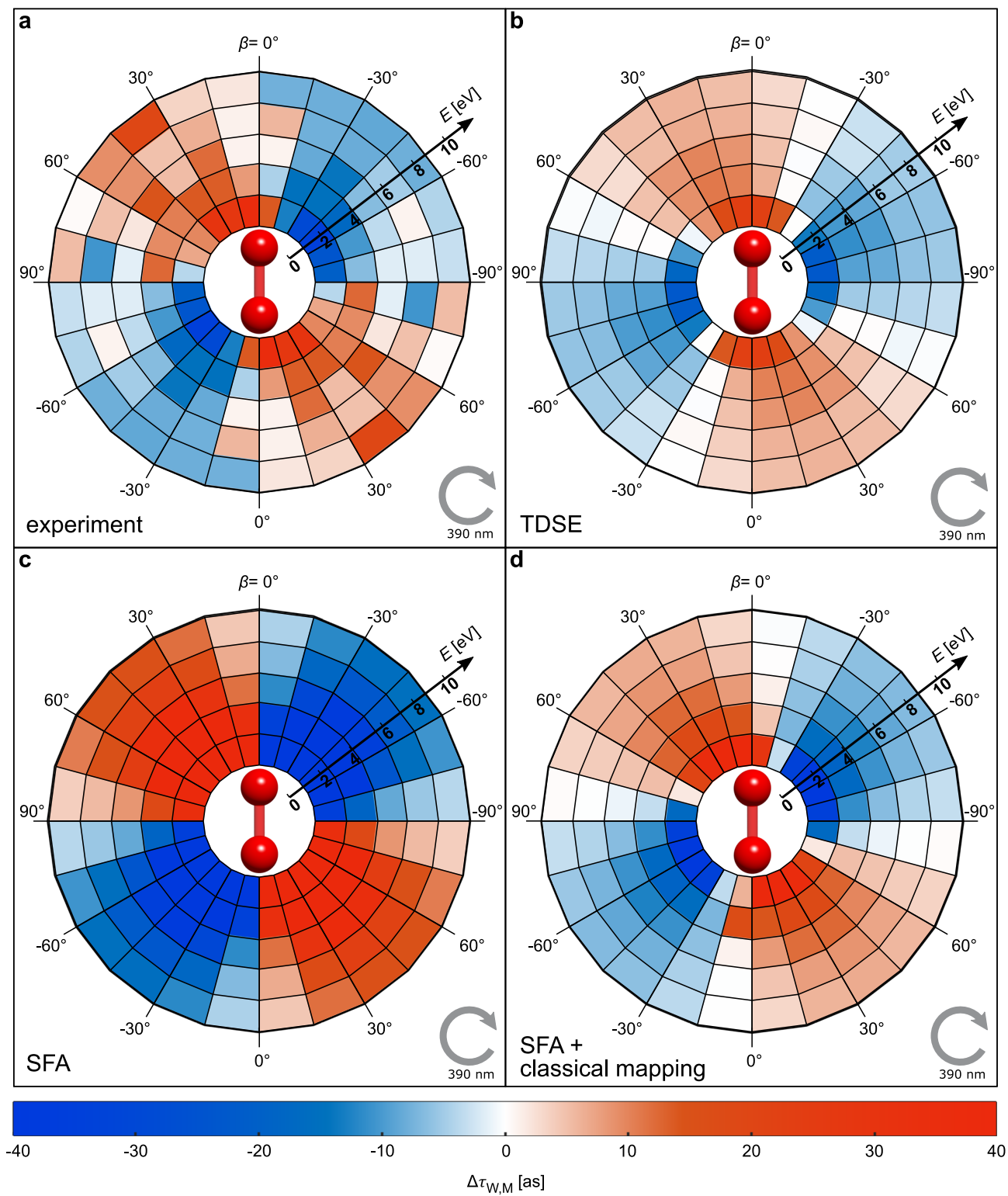

**Fig. 4 Changes of the Wigner time delay in the molecular frame. a** Experimentally retrieved changes of the Wigner time delay, $\Delta\tau_{W,M}$, as a function of the electron energy and $\beta$ (relative angle between electron momentum vector $\mathbf{p}_{elec}$ and the molecular axis of $H_2$ in the polarization plane). **b** $\Delta\tau_{W,M}$ from solving the time-dependent Schrödinger equation (TDSE). **c** $\Delta\tau_{W,M}$ from strong-field approximation (SFA). **d** $\Delta\tau_{W,M}$ from our classical-trajectory mapping of the initial momenta to the final momenta (including a long-range $1/r$ potential) with the initial phase gradient in momentum space at the tunnel exit from SFA. The molecular axis is shown schematically in all panels. The gray arrow indicates the light's helicity.

This agreement of SFA and experiment contrasts with the slight offset in $\beta$ that is observed comparing experiment and TDSE and warrants further research. One possible reason for the slight offset in $\beta$ could be a contribution to $\Delta\tau_{W,M}$ that is due to the long-ranging tail of the molecular potential that is included in our TDSE simulation while for the classical mapping that is used for Fig. 4a, d only an angularly independent $1/r$ potential is used (see "Methods"). Another reason for the slight offset in $\beta$ might be that the value $\Delta\phi'_{init}$ from SFA and experiment only accounts for the phase gradient that is perpendicular to the tunneling direction but the gradient along the tunneling direction is not included which might also slightly affect $\Delta\tau_{W,M}$. A third reason could be the choice of the pulse duration that is used for the TDSE calculation (see "Methods"). We expect that the first two possible contributions would be minimized for longer tunnel exits. Hence, we expect that there is a regime of long wavelengths and low intensities where the agreement of TDSE and experiment would be further improved.

## Discussion

We have measured the angular dependence of the Wigner time delay $\Delta\tau_{W,M}$ in strong-field ionization of the $H_2$ molecule. The main contribution to $\Delta\tau_{W,M}$ results from a microscopic position offset of the initial electronic wave function upon tunneling, which is determined by the spatial dimensions of molecular hydrogen. We also measure a slight modification of $\Delta\tau_{W,M}$ by the nonadiabaticity of the tunneling process, and the long-range part of the molecular potential. The simple orbital shape of the hydrogen molecule allows for an intuitive interpretation of the experimental results. A numerical simulation of the TDSE serves as high-level validation of our results, where $\Delta\tau_{W,M}$ is calculated directly from the phase of the continuum wave function. The quantitative agreement of experiment and theory is an important benchmark, showing that spatial information regarding the molecular orbital is accessible in strong-field ionization. Our findings pave the way towards a new class of experiments that can measure sub-Ångstrom position offsets and related changes of the Wigner time delay for electrons that tunnel from atoms and molecules.

## Methods

**Laser setup and gas target preparation**. The CoRTC pulses are generated in an interferometric setup based on a 200-μm β-barium borate crystal to double the frequency of laser pulses with a central wavelength of 780 nm (KMLabs Dragon, 40-fs FWHM, 8 kHz). The optical setup is the same as in refs. [38,39]. The light is focused by a spherical mirror ($f = 80$ mm) onto a cold supersonic jet of $H_2$. Intensity calibration is done as in ref. [37] and yields intensities of $I_{390\,nm} = 9.4 \times 10^{13}$ W/cm$^2$, $I_{780\,nm} = 5.5 \times 10^{11}$ W/cm$^2$ (corresponding to peak electric fields of $E_{390\,nm} = 0.037$ a.u. and $E_{780\,nm} = 0.0028$ a.u.). This leads to a Keldysh parameter that is approximately $\gamma_{390\,nm} = \frac{\omega_{390\,nm}\sqrt{2I_p}}{E_{390\,nm}} = 3.36$ with the light's angular frequency $\omega_{390\,nm} = 0.117$ a.u. and the ionization potential $I_p = 15.43$ eV of $H_2$. This is the regime of nonadiabatic tunneling[31,36,37]. The uncertainty of the absolute intensity is estimated to be 20%. The relative phase between the two single-color laser pulses was actively scanned during the measurement, long-term drifts were compensated in the offline analysis by a rotation of the momentum distribution in the polarization plane as in ref. [37]. The absolute orientation of the laser electric field in the plane of polarization, which is shown in Fig. 2a, is not known from the experiment and is estimated from the angular distribution of the intensity envelope that is presented in Fig. S1(b). The absolute orientation of the laser electric field does not affect the obtained values for $\Delta\alpha$ or $\Delta\tau_{Wigner}$ in a good approximation[13]. The supersonic gas jet was created by expanding hydrogen gas at a pressure of 2.5 bar through a 30 μm diameter nozzle into the vacuum.

**Particle detection and analysis**. We use a COLTRIMS reaction microscope[32] with an electron acceleration length of 390 mm and an ion acceleration length of 66 mm. An electric field of 10.9 V/cm and a magnetic field of 8.1 Gauss are applied to guide the charged particles to the respective detectors. The detectors are each comprised of a double-stack of micro-channel plates (diameter electron detector: 120 mm; ion detector: 80 mm) followed by position- and time-sensitive hexagonal delay-line anodes[40]. Position and time-of-flight information is used to calculate the three-dimensional momentum vectors of all charged particles in coincidence. Only events corresponding to the main dissociation channel (via the $\sigma_{gerade}$ state) have been selected by gating on a kinetic energy release between 0.8 and 2.2 eV[41]. The measured photoelectron momentum distribution in Fig. 2b can be considered as the product of the interference pattern we are after and an intensity envelope. The envelope is determined by the tunneling probability which is a function of initial momentum and the time-dependent laser electric field[42]. As shown in Fig. S1b, the envelope can be obtained from the full electron momentum spectrum presented in Fig. S1a by filtering higher Fourier components along the radial direction while maintaining the integral for each angle (see ref. [43]). Division of the full distribution by the extracted envelope leads to the normalized spectrum shown in Fig. S1c which is then used for further analysis. This procedure is conducted independently for each value of $\beta$. For each peak in the radial direction in Fig. S1c, the most probable electron emission angle $\alpha$ was determined from the onefold symmetric angular distribution (as described in ref. [13]). The values for $\alpha(\beta)$ are depicted in Fig. S1d–f. For each energy peak a reference value $\alpha_{mean}$ (mean of all values of $\alpha$ for this energy peak) has been calculated (indicated in Fig. S1e, f by horizontal lines). These reference values have been subtracted for each energy peak independently in order to obtain the angles $\Delta\alpha(\beta) = \alpha(\beta) - \alpha_{mean}$ that are depicted in Fig. 2c. The value of $\beta$ is between −90° and +90° which is assured in the analysis by taking the twofold symmetry of $H_2$ into account and subtracting or adding 180° to $\beta$, if necessary. As can be seen from the unsymmetrized data shown in Fig. S1, the assumption of this twofold symmetry is well supported by the experimental data (but in general, slight asymmetries can occur[44]). The situation depicted in Fig. 2a corresponds to a positive value of $\beta$. Error estimation has been done throughout this work by dividing the entire data set into three subsets which were analyzed separately. The error was calculated as the standard deviations from the three independent results. Thus, error bars show statistical errors only.

**Modeling of the spatial displacements perpendicular to the tunnel exit within the PFT model**. We use the LCAO single electron wave function for an internuclear distance $R = 0.74$ Å. We calculate the tunnel exit to be at a radius of $r_e = \frac{I_p}{E_{mean}} = 15.75$ a.u. from the origin of the coordinate system, for the peak electric field in the experiment and the ionization potential of $H_2$ ($I_p = 15.43$ eV). Here, $E_{mean} = 0.037$ a.u. is the average of the absolute value of the laser electric field for one cycle of the CoRTC field that is used in our experiment. The coherent superposition of the two 1s orbitals is represented by the wave function in position space:

$$\psi(x,y) \sim \exp\left(-\sqrt{x^2 + (y - R/2)^2}\right) + \exp\left(-\sqrt{x^2 + (y + R/2)^2}\right) \quad (3)$$

This expression is evaluated at the tunnel exit's position along a straight line $s_\perp$ that is a tangent to the circle with the radius $r_e$ and is perpendicular to the laser electric field at the instant of tunneling (see Fig. 1a). This results in a one-dimensional subset in position space $\psi_\perp(s_\perp)$ that models the wave function in position space perpendicular to the laser electric field at the instant at tunneling:

$$\psi_\perp(s_\perp) \sim \psi(x(s_\perp), y(s_\perp)) \quad (4)$$

Here, $x(s_\perp) = r_0(s_\perp) \cdot \sin(\eta(s_\perp))$ and $y(s_\perp) = r_0(s_\perp) \cdot \cos(\eta(s_\perp))$ with $r_0(s_\perp) = \sqrt{r_e^2 + s_\perp^2}$ and $\eta(s_\perp) = \gamma - \tan^{-1}\left(\frac{s_\perp}{r_e}\right)$, $\gamma$ denotes the relative angle of the laser electric field at the instant of tunneling with respect to the molecular axis.

Fourier transformation of $\psi_\perp(s_\perp)$ determines the phase of the initial electron wave packet ($\phi_{init}(p_{initial})$) in the direction that is perpendicular to the tunneling direction[15,16]. Also, note that the tunnel exit radius in Fig. 1a is chosen to be 1.3 a. u. for illustrational purposes only. In order to calculate the results in Fig. 1b the real value for the tunnel exit radius of $r_e = 15.75$ a.u. is used.

**Calculation of the changes of the Wigner time delay as a function of $\beta$ and the electron energy using the PFT model**. In order to calculate the Wigner time delay independent of the experiment (results are shown in Fig. 1d and Fig. S4e) we use the complex-valued initial wave function in momentum representation upon

tunnel ionization of $H_2$ (Eq. 17 from ref. [16]) and evaluate the phase gradient $\phi'_{\text{init}}$ at the most probable initial momentum for adiabatic tunneling ($p_{\text{initial}} = 0$ a.u., ref. [45]). This phase gradient is given by $\phi'_{\text{init}}(p_{\text{initial}}) = \frac{d\phi_{\text{init}}(p_{\text{initial}})}{dp_{\text{initial}}}$. We define the initial momentum perpendicular to the tunnel exit $p_{\text{initial}}$ by:

$$p_{\text{initial}} = \begin{cases} |\mathbf{p}_{\text{init}}|, & \text{if } \mathbf{p}_{\text{init}} \cdot \mathbf{A}(t_0) < 0 \\ -|\mathbf{p}_{\text{init}}|, & \text{if } \mathbf{p}_{\text{init}} \cdot \mathbf{A}(t_0) > 0 \end{cases} \quad (5)$$

Here, we use $\phi'_{\text{init}} = 0$ rad/a.u. as a reference and therefore use the identity $\phi'_{\text{init}} = \Delta\phi'_{\text{init}}$ without loss of generality. For typical tunneling geometries, the phase of the initial wave function in momentum representation is almost linear. Thus, the value of $\Delta\phi'_{\text{init}}$ is very similar for all initial momenta that correspond to nonvanishing amplitudes (see ref. [17] for examples).

After the tunneling step, the initial wave packet is accelerated in the electric field and subsequently mapped to the final momentum distribution. If we neglect the ionic potential and model the acceleration classically, the electric field at the instant of tunneling is perpendicular to the final momentum $\mathbf{p}_{\text{elec}}$ ($\gamma = \beta + 90°$) and the relation $\mathbf{p}_{\text{elec}} = \mathbf{p}_{\text{initial}} + e|\mathbf{A}(t_0)|$ holds. As a result, the orientation-dependent part of the Wigner time is given by $\Delta\tau_{\text{W,M}} = \hbar \frac{m_e}{|\mathbf{p}_{\text{elec}}|} \phi'_{\text{init}}$ (see Eq. 11 from ref. [13]). The procedure to obtain $\Delta\phi'_{\text{init}}$ and $\Delta\tau_{\text{W,M}}$ is carried out for each $\beta$ separately using that Eq. 17 from ref. [16] depends on the molecular orientation.

It should be noted that in the PFT model the wave packet after tunneling is Fourier-transform-limited and the peak of the wave function in position space travels with the group velocity of the wave packet. If the wave packet after tunneling is not Fourier-transform-limited, the shape of the wave function in position space can have a nonintuitive time-dependence. However, in any case, $\Delta\tau_{\text{W,M}}$ is a well-defined quantity that characterizes the change of the spectral phase of the photoelectron's wave packet[46].

Within the PFT model, the measured rotational offsets in final electron momentum space (measured as $\Delta\alpha$ for every energy peak and every $\beta$) can be translated to (i) changes of the phase gradient ($\Delta\phi'_{\text{init}}$ as a function of $E$ and $\beta$), (ii) changes of the Wigner time delay ($\Delta\tau_{\text{W,M}}$ as a function of $E$ and $\beta$), and (iii) position offsets that are parallel to the final electron momentum ($\Delta s_\perp$ as a function of $\beta$ since $\phi_{\text{init}}$ is linear in $p_{\text{initial}}$). All three quantities can be viewed as different ways to model the same physical reality.

**Extraction of $\Delta\phi'_{\text{init}}$ from the experimentally accessible quantity $\Delta\alpha$.** The changes of the phase gradient in momentum space at the tunnel exit $\Delta\phi'_{\text{init}}$ (Fig. 3a) are calculated from the data that is shown in Fig. 2c. The theoretical framework that is used for this purpose is described in detail in ref. [13] and summarized in the following. In ref. [13] a semiclassical, trajectory-based model (SCTS model) is introduced that reproduces the alternating half-ring (AHR) pattern for CoRTC fields. It is found that a change of $\Delta\alpha$ reflects a change of the phase gradient $\Delta\phi'_{\text{init}}$ and that this relationship holds for every energy peak individually (see Fig. 10d in ref. [13]). We perform our SCTS simulation as in ref. [29] but use $E_{390} = 0.037$ a.u., $E_{780} = 0.0028$ a.u., a two-cycle flat-top light pulse, restrict the electron release time to these two light cycles, use $E_{\text{SFA}} = 0.037$ (see Eq. A2 from ref. [29]) and added a phase in momentum space for each trajectory which is defined by $\phi_{\text{init}}(p_{\text{initial}}) = p_{\text{initial}} \Delta\phi'_{\text{init}}(p_{\text{initial}})$.

Analyzing how an assumed value of $\Delta\phi'_{\text{init}}$ leads to changes of $\alpha$ allows for the creation of a look-up-table that provides $\Delta\phi'_{\text{init}}$ for a measured combination of $E$ and $\Delta\alpha$. The result is shown in Fig. S2 and this look-up-table is independent of $\beta$ because in the SCTS model only the asymptotic $1/r$ part of the long-range Coulombic potential is included. For a given value of $E$ and small values of $\Delta\alpha$, there is an almost linear relation between $\Delta\alpha$ and $\Delta\phi'_{\text{init}}$.

**Classical mapping of $\beta$ and $p_{\text{elec}}$ to $\gamma$ and $p_{\text{initial}}$ including a $1/r$ potential after tunneling.** For Fig. 3a, Fig. 4a, d we use a classical mapping that includes a $1/r$ potential after tunneling. To this end, we use classical trajectories and solve Newton's equations after tunneling in the presence of the laser electric field and a $1/r$ potential. We use the same SCTS simulation as for Fig. S3 but include only the light pulse with a central wavelength of 390 nm and neglect the phase of the trajectories. As a result, the classical trajectories provide an unambiguous link between $p_{\text{elec}}$ and $p_{\text{initial}}$ as shown in Fig. S3a.

Moreover, Coulomb interaction after tunneling[18,19] affects the mapping of $\beta$ to $\gamma$. The angular offset, which is due to Coulomb interaction after tunneling, is represented by the value $\kappa$, which is defined as the offset angle between $-\mathbf{A}(t_0)$ and $\mathbf{p}_{\text{elec}}$ in the polarization plane, and which is shown as a function of $p_{\text{initial}}$ in Fig. S3b. Hence, the measured values of $\beta$ are mapped to the values of $\gamma$ using $\gamma = \beta + 90° - \kappa$.

We note that our experiment gives access to the phase gradient $\Delta\phi'_{\text{init}}$ using our CoRTC field. As described above, in our CoRTC fields there are two slightly different values of $p_{\text{initial}}$ for each final electron momentum. The classical mapping functions are an approximation for which we neglect the weak field at a central wavelength of 780 nm in order to obtain unambiguous mapping functions.

Please note that for the results for the PFT model (Fig. 1d) and SFA (Fig. 3c) we use classical potential-free trajectories for the classical mapping of $\beta$ and $p_{\text{elec}}$ to $\gamma$ and $p_{\text{initial}}$ (see Eq. 9 and the dashed black lines in Fig. S3a, b).

**Determine the changes of the Wigner time delay $\Delta\tau_{\text{W,M}}$ in the experiment.** In order to obtain the experimental result for $\Delta\tau_{\text{W,M}}(E,\beta)$ that is shown in Fig. 4a, the change of the Wigner time delay is expressed by:

$$\Delta\tau_{\text{W,M}}(E,\beta) = \frac{\hbar m_e}{p_{\text{elec}}} \Delta\phi'_{\text{elec}}(p_{\text{elec}},\beta) \approx \frac{\hbar m_e}{p_{\text{elec}}} \Delta\phi'_{\text{init}}(p_{\text{init}}(p_{\text{elec}}),\gamma(\beta)) \frac{dp_{\text{init}}}{dp_{\text{elec}}} \quad (6)$$

The definitions $\phi'_{\text{init}}(p_{\text{initial}},\gamma) = \frac{d\phi_{\text{init}}(p_{\text{initial}},\gamma)}{dp_{\text{initial}}}$ and $\phi'_{\text{elec}}(p_{\text{elec}},\beta) = \frac{d\phi_{\text{elec}}(p_{\text{elec}},\beta)}{dp_{\text{elec}}}$ are used. $\Delta\phi'_{\text{init}}(p_{\text{initial}},\gamma)$ is known from Fig. 3a. The same classical simulation as for Fig. S3a, b is used to obtain the classical mapping functions $p_{\text{initial}}(p_{\text{elec}})$ and $\gamma(\beta)$. Evaluating Eq. 6 allows one to calculate $\Delta\tau_{\text{W,M}}$ for all measured combinations of $E$ and $\beta$. The result is shown in Fig. 4a.

The findings regarding $\Delta\phi'_{\text{init}}$ and $\Delta\tau_{\text{W,M}}$ are valid for the strong field ionization by circularly polarized light. For differently polarized lights fields, the lack of rotational symmetry implies that the initial phase gradient must be retrieved in a more complex manner and that sub-cycle interferences might affect the obtained phase gradient in the electron continuum[28,47].

It is important to note that the Wigner time delay and the most probable electron release time[18,48–50] are different quantities. In typical experiments that use the attoclock-setup the square of the electronic wave function in final momentum, space is measured[18,19,50–52]. If no interference patterns are measured, the result is often interpreted to reflect the amplitudes of the electronic wave packet in the continuum. Assuming that this electronic wave packet is born at the tunnel exit and streaked to its final momentum in the presence of an atomic or molecular potential allows one to infer the amplitudes of the wave packet in momentum space at the tunnel exit.

Our experimentally obtained phase gradients are independent of the amplitudes in the final momentum space to a good approximation for two reasons. First, for the retrieval of $\Delta\alpha(\beta)$, the modulation of the ionization rate as a function of the laser electric field's magnitude is removed by normalization (see Fig. S1). Second, we analyze the interference patterns for a given value of $\beta$ which makes our analysis insensitive to the dependence of the ionization rate on the tunneling direction in the molecular frame[48,53]. Therefore, HASE allows for the measurement of the phase of the initial wave packet in momentum space at the tunnel exit, just as the attoclock setup can be used to measure the amplitude of the initial wave packet in momentum space at the tunnel exit.

**TDSE simulations.** We perform numerical simulations of the time-dependent Schrödinger equation (TDSE), $i\partial_t\psi(\mathbf{r},t) = H(t)\psi(\mathbf{r},t)$, with a Hamiltonian $H(t) = \frac{1}{2}(\mathbf{p} + \mathbf{A}(t))^2 + V(\mathbf{r})$ in single-active electron approximation. This ab initio approach allows for the calculation of the orientation dependence of the electron position. In two dimensions, we consider a model $H_2$ molecule (with fixed nuclei) described by a soft-core potential

$$V(\mathbf{r}) = -\sum_{j=1,2} \frac{Z_{\text{eff}}}{\sqrt{\mathbf{r}_j^2 + \epsilon}}, \quad (7)$$

where $\mathbf{r}_j = \mathbf{r} - \mathbf{R}_j$ with $\mathbf{R}_1$ and $\mathbf{R}_2$ being the positions of the nuclei at equilibrium distance $R_{\text{eq}} \approx 1.398$ a.u. With an effective charge $Z_{\text{eff}} = 0.5$ a.u. and a soft-core parameter $\epsilon = 0.265$ a.u., we reproduce the ionization potential of real $H_2$. In order to suppress the influence of inter-cycle interferences, we use circularly polarized laser pulses with a 3-cycle $\sin^4$-envelope for the vector potential $\mathbf{A}(t)$. The angular frequency is $\omega_{390\text{nm}}$ (corresponding to a wavelength of $\lambda = 390$ nm) and the maximal absolute value of the vector potential is 0.317 a.u. We find that there is only a small difference in the orientation dependence of the phase gradient in final electron momentum space (for momenta at the positions of the ATI rings) if we use longer light pulses (11 optical cycles).

In order to solve the TDSE we divide the configuration space in an inner region and an asymptotic outer region (see refs. [54,55] for details). In the inner part we solve the TDSE using the split-operator method on a Cartesian grid with 4096 points in each dimension, a grid spacing of 0.25 a.u. and a time step of 0.02 a.u. In the asymptotic region the interaction of the electron with the ionic potential is neglected. Hence, this wave function is represented in momentum space at all times and it is propagated until a final time $t_f$ by means of Volkov states. The inner region has an absorbing boundary covering a distance of 50 a.u. from the edge of the grid, i.e., the absorber is present at $|\mathbf{r}| > 462$ a.u. The absorbed parts of the inner wave function are not discarded but added coherently to the wave function of the outer region. The momentum-space wave function $\psi(\mathbf{p})$ of the (outgoing) ionized wave packet is thus collected in the outer part.

To study the changes of the Wigner time delay as a function of the molecular orientation, we evaluate the energy derivative along the direction that corresponds to the maximum of the negative vector potential and refer to this as $\frac{d\phi_{\text{TDSE}}(E,\beta)}{dE}$. Subsequently, we subtract the mean value over all molecular orientations (here 24 different calculations) and obtain $\Delta\tau_{\text{W,M}}(E,\beta) = \hbar \frac{d\phi_{\text{TDSE}}(E,\beta)}{dE} - b$, where $b$ is the mean of $\hbar \frac{d\phi_{\text{TDSE}}(E,\beta)}{dE}$ over all $\beta$. The results are shown in Fig. 4b.

**SFA simulations.** The SFA offers an approximate way to describe the quantum dynamics in strong laser fields. Here, the influence of the ionic potential on the outgoing electron wave packet is neglected, but nonadiabatic effects such as an initial momentum are included inherently[31,36,37]. In length-gauge, the momentum amplitude of the outgoing wave packet at a final time $t_f$ can be written in an integral form[56–58]

$$\widetilde{\psi}_{\text{SFA}}(\mathbf{p}, t_f) = -i \int_0^{t_f} dt' \, \mathbf{E}(t') \cdot \mathbf{u}(\mathbf{p} + \mathbf{A}(t')) \, e^{iS_{\text{SFA}}(\mathbf{p},t')} \quad (8)$$

with the action $S_{\text{SFA}}(\mathbf{p}, t') = I_p t' + \int_{t'}^{t_f} \frac{1}{2}[\mathbf{p} + \mathbf{A}(\tau)]^2 d\tau$. The dependence on the molecular orientation enters only through the dipole matrix element $\mathbf{u}(\mathbf{v}) = \langle \mathbf{v} | \mathbf{r} | \psi_0 \rangle$, i.e., the initial state of the electron $\psi_0$ of Eq. 3.

In order to evaluate the integral of Eq. 8 numerically, we use the same short laser pulse as for the TDSE calculations and follow the same procedure in order to calculate $\Delta \tau_{\text{W,M}}(E, \beta)$. The result is shown in Fig. 4c.

The phase gradient as a function of the tunneling direction in the molecular frame and the initial momentum $\Delta \phi'(p_{\text{initial}}, \gamma) = \frac{d\phi_{\text{SFA}}(p_{\text{initial}}, \gamma)}{dp_{\text{initial}}}$ is obtained by mapping the final momenta $\mathbf{p}_{\text{elec}}$ to the initial momenta $\mathbf{p}_{\text{init}}$ using classical potential-free trajectories with:

$$\mathbf{p}_{\text{elec}} = -e\mathbf{A}(t_0) + \mathbf{p}_{\text{init}} \text{ and } \gamma = \beta + 90° \quad (9)$$

Here, $\mathbf{A}(t_0)$ is the vector potential at the time $t_0$. This leads to the result that is shown in Fig. 3b. The result shown in Fig. 4d has been obtained from the data shown in Fig. 3b in full analogy to the way the experimental data was processed to obtain Fig. 4a from Fig. 3a.

### Data availability
The data that support the plots within this Article is available from the corresponding authors upon reasonable request.

### Code availability
The code that supports the plots within this Article is available from the corresponding authors upon reasonable request.

### Acknowledgements
This work was funded by the German Research Foundation (DFG) through priority programme SPP 1840 QUTIF.

### Author contributions
D.T., K.F., N.A., A.G., S.G., M.S.S., L. Ph. H.S., T.J., R.D., M.K., and S.E. contributed to the experiment. S.B., D.T., R.D, and S.E. contributed to the theoretical results. D.T., R.D., and S.E. performed the analysis of the experimental data. All authors contributed to the paper.

### Funding
Open Access funding enabled and organized by Projekt DEAL.

### Competing interests
The authors declare no competing interests.

### Additional information
**Supplementary information** The online version contains supplementary material available at https://doi.org/10.1038/s41467-021-21845-6.

**Correspondence** and requests for materials should be addressed to D.T. or S.E.

**Peer review information** *Nature Communications* thanks Matthias Kling and the other, anonymous, reviewer(s) for their contribution to the peer review of this work.

**Reprints and permission information** is available at http://www.nature.com/reprints

**Publisher's note** Springer Nature remains neutral with regard to jurisdictional claims in published maps and institutional affiliations.

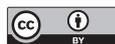